\documentclass[graybox,vecphys,url]{svmult}
\usepackage{mathptmx} 
\usepackage{helvet} 
\usepackage{courier} 
\usepackage{type1cm} 
\usepackage{makeidx} 
\usepackage{graphicx} 
\usepackage{epsfig,rotating}

\usepackage{multicol} 
\usepackage[bottom]{footmisc}

\def\cm{cm$^{-1}\,$}
\def\nNH2{$\nu_{\rm NH_{2}}\,$}
\def\dNH2{$\delta_{\rm NH_{2}}\,$}
\def\nCnHB{$\nu_{\rm C2=O2}\,$}
\def\nCHB{$\nu_{\rm C4=O4}\,$}
\def\dH2O{$\delta_{\rm H_{2}O}\,$}
\def\nH2O{$\nu_{\rm H_{2}O}\,$}

\begin{document}

\title*{Anharmonic Vibrational Dynamics of DNA Oligomers}
\author{O. K\"uhn, N. Do\v{s}li\'{c}, G. M. Krishnan, H. Fidder, K. Heyne}
\authorrunning{K\"uhn et al.}
\institute{O. K\"uhn \at Institut f\"ur Physik, Universit\"at Rostock, Universit\"atsplatz 3, 18051 Rostock, Germany\\ \email{oliver.kuehn@uni-rostock.de}
\and N. Do\v{s}li\'{c} \at Department of Physical Chemistry, Rudjer Bo\v{s}kovi\'{c} Institute, 10000 Zagreb, Croatia
\and G. M. Krishnan \at Institut f\"ur Chemie und Biochemie, Freie Universit\"at Berlin, Takustr. 3, 14195 Berlin, Germany
\and H. Fidder, K. Heyne \at Institut f\"ur Physik, Freie Universit\"at Berlin, Arnimallee, 14195 Berlin, Germany\\ \email{heyne@physik.fu-berlin.de}}
%
%
\maketitle

\abstract{
Combining two-color infared pump-probe spectroscopy and anharmonic force field calculations we characterize the anharmonic coupling patterns between fingerprint modes and the hydrogen-bonded symmetric \nNH2 stretching vibration in adenine-thymine  dA$_{20}$-dT$_{20}$  DNA oligomers.  Specifically, it is shown that the anharmonic coupling between the \dNH2 bending and the \nCHB stretching vibration, both absorbing around 1665 \cm, can be used to assign the \nNH2 fundamental transition at 3215 \cm despite the broad background absorption of water.
}
%
\section{Introduction}
\label{DNA:sec:into}
%
Vibrational energy redistribution and relaxation in complex systems
depends on the network of anharmonically coupled vibrational states
subject to fluctuations due to the interaction with some environment
\cite{may04}. Focussing on hydrogen-bonded systems there is
considerable evidence that the time scales for relaxation can be in
the subpicosecond range pointing to a rather strong interaction, e.g., of the excited
stretching vibration with other hydrogen bond (HB) related modes
such as the bending and the low-frequency HB distance vibration as
well as with the solvent \cite{nibbering04:1887,heyne04:6083,nibbering07:619,giese06:211,marechal07}.

One of the most prominent hydrogen-bonded systems is DNA.
Despite numerous experimental and theoretical investigations on
vibrational spectra of nucleic acid bases \cite{tsuboi69:45,clark85,lettelier87:663,florian94:1457,florian95:421,ouali97:4816,shishkin99:15}, information on inter- and intramolecular interactions in base pairs and DNA oligomers is
still limited \cite{howard65:801,spirko97:1472,sponer97:76,brandl99:103,sponer01:43,nir02:740,krummel03:9165,woutersen04:5381,lee06:114508,lee06:114509,lee06:114510,lee07:145102}.  A recent example is the work on single
adenine-uracil (AU) base pairs in the Watson-Crick geometry in
solution, which showed an enhancement of vibrational energy
relaxation of the NH stretching vibration by a factor of three as compared to the isolated uracil base \cite{woutersen04:5381}.

\begin{figure}[t]
\begin{center}
\includegraphics[scale=.8]{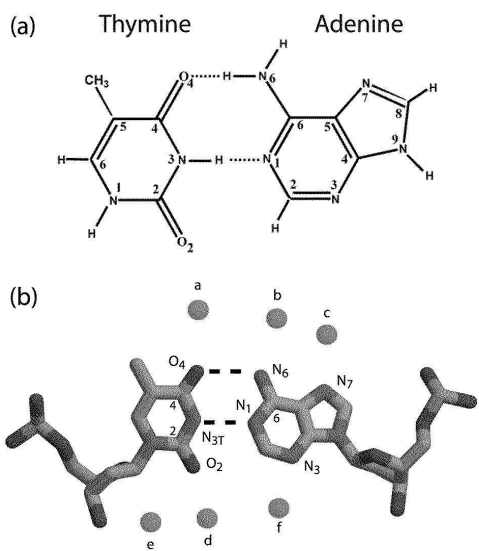}
\caption{(a) Scheme of the AT DNA Watson-Crick configuration. b)
Structure of a single AT DNA base pair in the Watson-Crick
configuration of a dodecamer (taken from 428d.pdb \cite{shui98:16877}). The oxygen
atoms of water molecules forming HBs to the nucleic acids
are presented as spheres. Distances of the oxygen atoms of water
molecules in the major groove a, b, and c to the O4, N6, and
N7 atoms are 2.93 {\AA}, 2.93 {\AA}, and 2.80 {\AA},
respectively. Water molecules d, e, and f of the minor groove have
distances of 2.88 {\AA}, 4.16 {\AA}, and 2.61 {\AA} to the O2 and
N3 atoms, respectively. Intrastrand distances of the O2,
O4, and N6 atoms to neighbooring thymine and adenine bases are
4.12 {\AA}, 3.53 {\AA}, and 3.34 {\AA}, respectively.}
\label{fig:DNAt}
\end{center}
\end{figure}

DNA oligomers adopt different types of conformations, both in gas
and condensed phases, such as the A, B, B', C, D, and Z form,
depending on water and salt concentration, type of cations, pH, and
base sequences \cite{clark85,ouali97:4816,lee07:145102,saenger84,parvathy02:1500,pattabiraman86:1603,pichler00:391}.  In the condensed phase the conformations of DNA oligomers are stabilized by water molecules that form water networks,
predominantly in the major and minor grooves, and near the phosphate groups of the backbone \cite{ouali97:4816}.  Among the different types of base sequences, adenine-thymine (AT) oligomers are special because they do not undergo transitions from the B to the A form upon reducing
the water content. Instead, AT oligomers adopt the B' form at low water concentrations, with 4 to 6 water molecules per base pair that
may be hydrogen-bonded to the oligomer \cite{ouali97:4816,taillandier87:3361,falk63:387,vargason01:7265}. In the B' form of
the AT DNA oligomer two HBs are formed in the Watson-Crick configuration, i.e.,  between oxygen (O4) of the thymine and the NH$_{2}$ group of the adenine (N6), and between the NH group of the thymine (N3T) and the nitrogen atom of the adenine (N1), see Fig. \ref{fig:DNAt}.

Vibrational modes expected to be strongly influenced by the hydrogen-bonding in the DNA helix are the carbonyl stretches \nCnHB at 1716 cm$^{-1}$  and \nCHB  at 1665 cm$^{-1}$  and the amine bending \dNH2 at 1665 cm$^{-1}$ \cite{tsuboi69:45,clark85,lettelier87:663,howard65:801,miles64:1104,adam86:3220,liquier91:177,sutherland57:446,ghomi90:691}.  Note, that in contrast to H$_2$O, in D$_2$O the $\delta_{\rm ND_{2}}\,$ vibration of adenine and the carbonyl vibrations of thymine are decoupled, due to
the frequency shift from \dNH2 to $\delta_{\rm ND_{2}}\,$ \cite{lee06:114508,lee06:114509}. The \dH2O vibration of water molecules in DNA samples typically absorbs in the same spectral region, i.e., around 1650 \cm \cite{tsuboi69:45,adam86:3220,benevides83:5747}. A
direct experimental assignment of \nNH2 and $\nu_{\rm NH}\,$ in AT DNA oligomers in the condensed phase is very difficult. Typically,
symmetric and antisymmetric \nNH2 stretching vibrations absorb around 3300 cm$^{-1}$ \cite{tsuboi69:45}. However, the spectral range from 3050 to 3600 cm$^{-1}$ is dominated by the strong absorption of the water OH stretching vibration. Reducing the water content of the DNA
oligomers does not solve this problem, because at extremely low water contents the DNA oligomers do not adopt a well defined
structure.

Ultrafast time-resolved infrared (IR) spectroscopy is ideally suited to address this issue as has been shown in studies of
inter- and intramolecular couplings and energy relaxation
dynamics  in various hydrogen-bonded systems \cite{nibbering04:1887,heyne04:6083,nibbering07:619,cowan05:199,heyne04:902}.  In this contribution we focus on shifts in oligomer vibrational modes induced by excitation of the \nCnHB or the \nCHB/ \dNH2 oligomer fingerprint vibration. These shifts originate from inter- and intramolecular couplings among different vibrational modes of the DNA oligomer and depend on the strength of the couplings as well as the energy mismatch between different
transitions. Related effects are particularly pronounced if overtones or
combination modes match a fundamental vibrational transition
(resonance enhancement). This already affects the linear absorption band
shape, but also the vibrational relaxation dynamics \cite{giese06:211}. A particular
strength of the ultrafast IR pump-probe spectroscopy is the
capability of uncovering vibrational spectral features not visible
in linear spectroscopy due to excessive solvent absorption. This is
demonstrated in the experiments presented here, where we excite
oligomer vibrations between 1600 and 1760 cm$^{-1}$ and probe for
the oligomer \nNH2 vibration in the region of 3050 - 3250 cm$^{-1}$, which is dominated by water absorption.

The experimental assignment of the adenine \nNH2 vibration and the coupling pattern across the HBs is supported by quantum
chemical calculations of anharmonic couplings which are used for
obtaining fundamental transitions frequencies for a set of  relevant modes
of a microsolvated gas phase AT model. 
 In principle
accurate theoretical modelling of the vibrational dynamics of DNA AT
base pairs requires taking into account several effects: (i) The
intermolecular double HB between adenine and thymine,
(ii) The interaction between different base pairs along the DNA
strand, (iii) The charges as well as the dynamics of the backbone,
and (iv) The influence of water molecules which may, for instance,
make a HB to the base pair. Here, we are aiming to obtain a
semiquantitative understanding of the transient band shifts, whereby
it is assumed that they are dominated by effect (i), that is, the
anharmonic coupling pattern due to the intermolecular HB.
The effect of (ii)-(iv) can be of static nature, e.g., changes in
the anharmonic frequencies and coupling constants, and also of
dynamic nature, e.g., fluctuation of the energy levels. However,
here we will only focus on the static influence of a well-defined
environment determined by microsolvation of the AT base pair by
several water molecules.

Isolated and microsolvated base pairs have been extensively studied theoretically, focussing
in particular on the stability of different isomers, see, e.g. the
work by Hobza and coworkers \cite{jurecka06:1985,kabelac05:12206,kabelac07:903} as well as by Fonseca-Guerra et al.
\cite{fonseca-guerra00:4117}. Although there is a number of reports on potential energy
surfaces of base pairs in harmonic approximation, there appear to be
only a few calculations addressing anharmonicity in the context of,
e.g., proton transfer \cite{florian94:1457,villani05:1,gorb04:10119}, the coupling to the intermolecular
HB vibration \cite{spirko97:1472} or the assignment of different gas phase
isomers \cite{krishnan07:132}. Most notable in this respect is the recent study of the
anharmonic spectrum of a guanine-cytosine pair \cite{brauer05:6974} as well as the
development of a vibrational exciton model to describe nonlinear IR spectra involving DNA fingerprint modes \cite{krummel03:9165,lee06:114508,lee06:114509,lee06:114510,lee07:145102,krummel06:13991}.

This Chapter is organized as follows. In the next Section we will first discuss the effect of solvating water molecules on the anharmonic IR spectra of an isolated AT pair. In this context we will scrutinize the applicability of a dual level approach which combines different quantum chemistry methods within a correlation expansion of the potential energy surface (PES). For the case of two water molecules we will present an analysis of the anharmonic coupling patterns between the \nCnHB, \nCHB and \dNH2 vibrations and the symmetric \nNH2 mode in Section \ref{sec:afc}. Section 3 gives details on the experimental setup and presents results of two-color pump-probe spectra. Finally, we give a comparison between theory and experiment in Section \ref{sec:disc} which leads us to the assignment of the \nNH2 fundamental transition.
%
\section{Microsolvated AT Base Pairs}
\subsection{Fundamental Transitions Using a Dual Level Approach}
In the following we present results on fundamental vibrational
transitions of isolated AT base pairs microsolvated with 1-4 water
molecules. The aim of this study is twofold: First to find out about
overall changes of IR transitions of base pair modes due to the
interaction with water molecules. And, second, to test the
performance of a dual level approach combining density functional
(DFT) and semiempirical PM3 data to expand the PES. Throughout we
will assume that the deviations from equilibrium structures are
small enough such as to allow the use of normal mode coordinates
$\mathbf{ Q}$ for spanning the PES, i.e. $V=V(\mathbf{ Q})$.

\begin{figure}[h]
\begin{center}
\includegraphics[scale=.6]{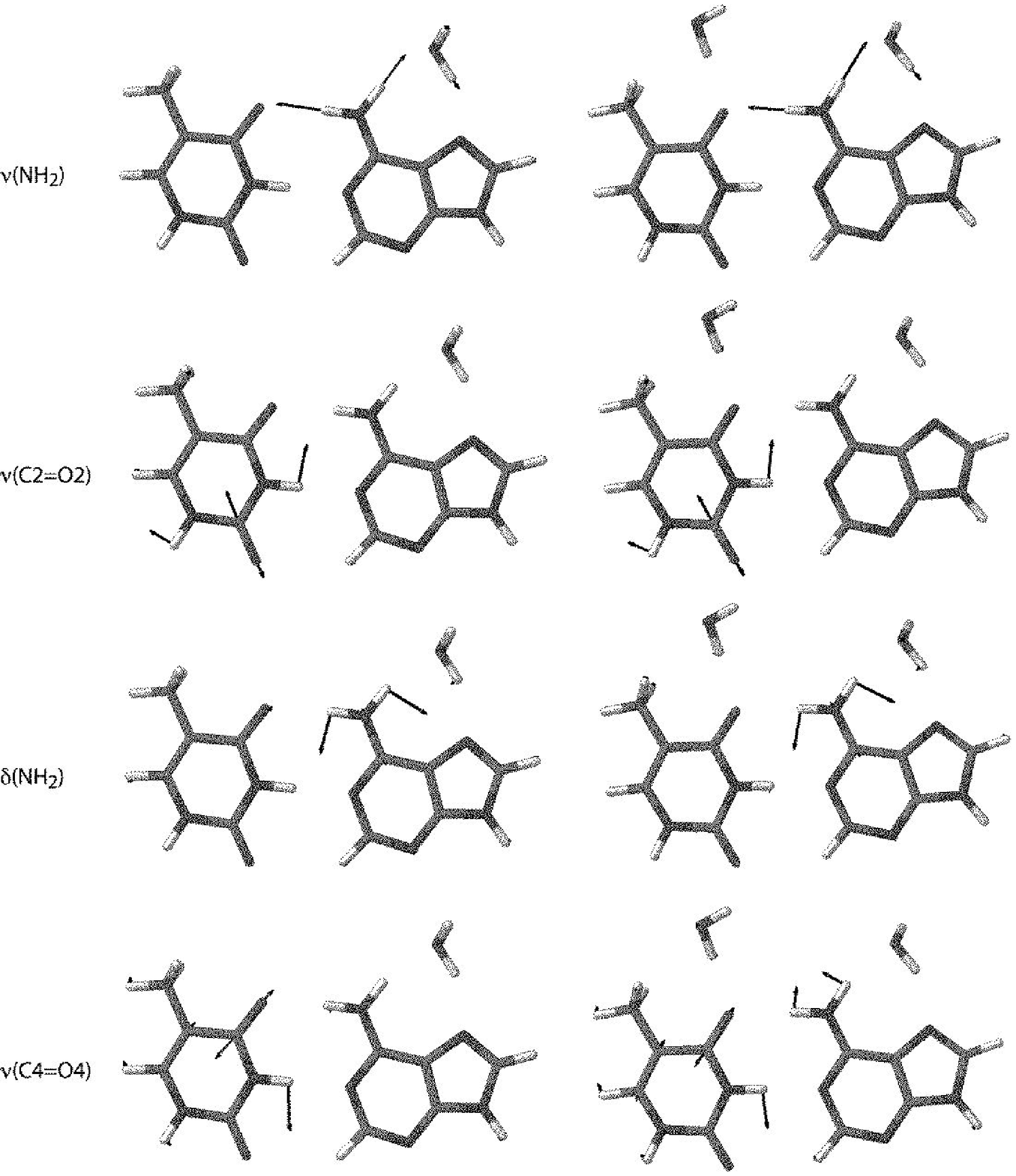}
\caption{Normal mode displacement vectors for the target modes as
calculated in harmonic approximation of the AT-(H$_{2}$O)$_{n=1,2}$
PES using DFT/B3LYP with a 6-31++G(d,p) basis set. }
\label{DNA:fig:NMs_1_2}
\end{center}
\end{figure}

\begin{figure}[h]
\begin{center}
\includegraphics[scale=.6]{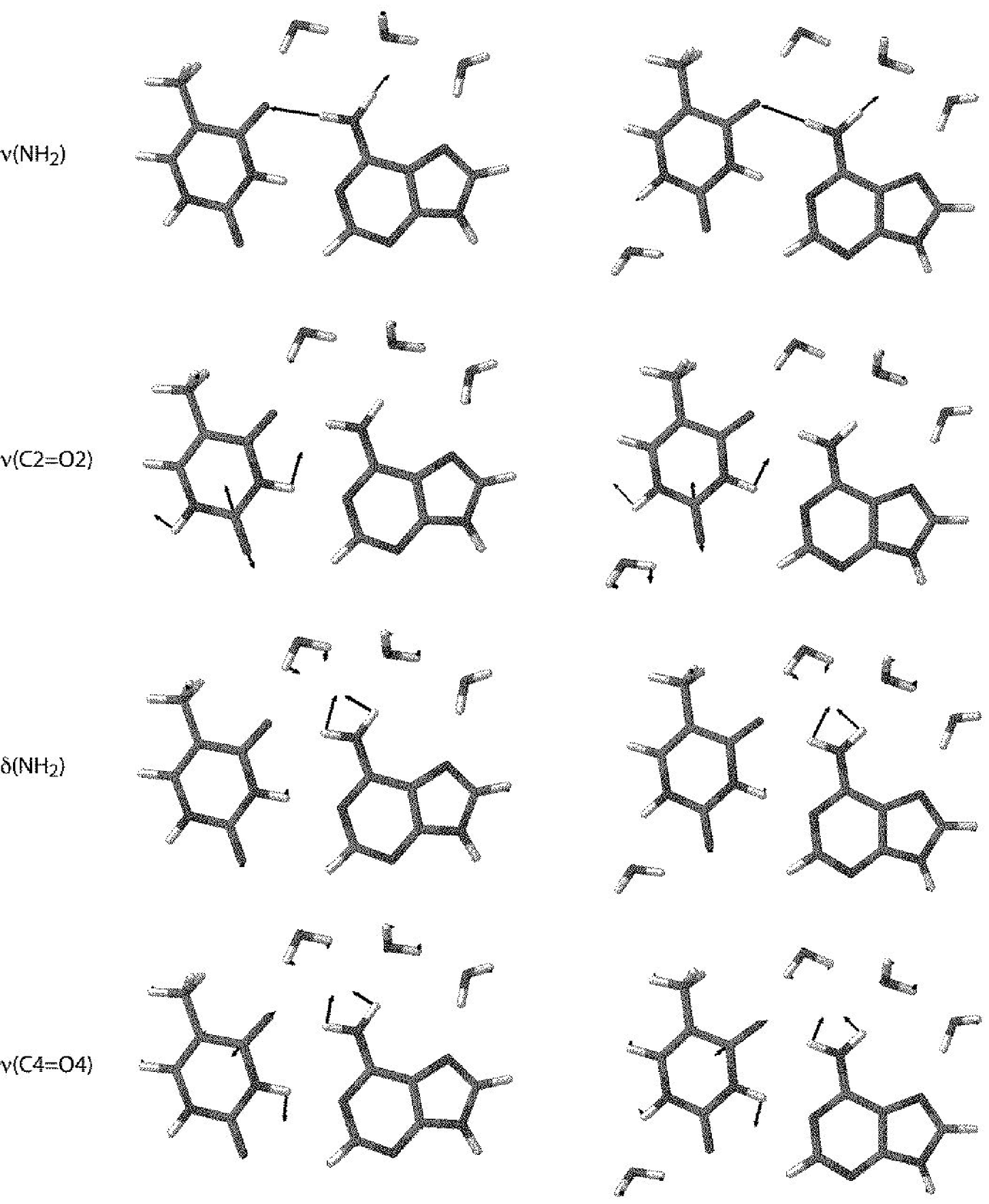}
\caption{Normal mode displacement vectors for the target modes as
calculated in harmonic approximation of the AT-(H$_{2}$O)$_{n=3,4}$
PES using DFT/B3LYP with a 6-31++G(d,p) basis set. }
\label{DNA:fig:NMs_3_4}
\end{center}
\end{figure}

Under the conditions of the experiment there are 4 to 6 water
molecules per AT pair which can form different HBs to the
base pair (see also Fig. \ref{fig:DNAt}). Our interest will be in the IR transitions of the NH$_2$
and C4=O4 groups such that water situated in the major groove shall
be of importance. However, for comparison we also consider a
structure where a water molecule is on the C2=O2 side. There are
several microsolvation studies which focussed on the effect of water
on base pair properties such as interaction energies or HB lengths
\cite{kabelac05:12206,kabelac07:903,fonseca-guerra00:4117,kumar05:3971}. To our
knowledge there is, however, no theoretical account on anharmonic IR
spectra of HB related modes. The four structures which
will be discussed in the following are shown in Figs.
\ref{DNA:fig:NMs_1_2} and \ref{DNA:fig:NMs_3_4}. They have been
obtained by geometry optimization at the DFT/B3LYP with a
6-31++G(d,p) level of theory using Gaussian 03 \cite{gaussian03}.
Notice that these are not necessarily the most stable structures at
this level of theory (see also discussion in Ref.
\cite{kabelac07:903}). Our choice has been biased by the requirement
that the water molecules should be close or even hydrogen-bonded to
the considered target modes. The latter are shown in terms of their
normal mode displacement vectors in Figs. \ref{DNA:fig:NMs_1_2} and
\ref{DNA:fig:NMs_3_4} as well. The respective harmonic frequencies
are compiled in Table \ref{tab1}.

In AT-H$_2$O, Fig. \ref{DNA:fig:NMs_1_2} (left column), the water
molecule is hydrogen-bonded between the adenine N6-H and the N7
sites. This causes the \nNH2 vibration to acquire some water
stretching character lowering its harmonic frequency. The \dNH2
vibration is only slightly mixed with some water motion and
essentially constrained so that its frequency is blue-shifted. The
next water molecule in AT-(H$_2$O)$_2$, Fig. \ref{DNA:fig:NMs_1_2}
(right column), makes a HB to the oxygen of C4=O4
lowering the \nCHB frequency slightly, but at the same time mixing
this vibration with \dNH2 type motions. For the case of three water
molecules, Fig. \ref{DNA:fig:NMs_3_4} (left column), there is the
possibility to form a hydrogen bonded water chain connecting the O4,
N6-H, and N7 sites. This reduces the mixing of the \nNH2 and water
motions, but the \dNH2 vibration contains a water bending now as
does the \nCHB mode. Adding another water at the C2=O2 site leads
as expected to a shift of the \nCnHB transition only, Fig.
\ref{DNA:fig:NMs_3_4} (right column). Overall we notice that the
presence of solvating water molecules has the strongest impact on
the \dNH2 and \nCHB vibrations, with the latter acquiring
substantial \dNH2 character.
\begin{table}[tdp]
\caption{Harmonic frequencies (in \cm) for the target modes of the
different model structures as obtained using DFT/B3LYP with a
6-31++G(d,p) basis set. The displacement vectors for the solvated
structures are shown in Figs. \ref{DNA:fig:NMs_1_2} and
\ref{DNA:fig:NMs_3_4}.}
\begin{center}
\begin{tabular}{lccccc}
\hline
modes & AT & AT-H$_2$O &AT-(H$_2$O)$_2$ &AT-(H$_2$O)$_3$ &AT-(H$_2$O)$_4$ \\
\hline
\nNH2 & 3410 & 3393 & 3401 & 3415 & 3410 \\
\nCnHB & 1797 & 1794 & 1799 & 1799 & 1776 \\
\dNH2 & 1689 & 1718 & 1720 & 1727 & 1727 \\
\nCHB & 1728 & 1731& 1714 & 1720 & 1720 \\
\hline
\end{tabular}
\end{center}
\label{tab1}
\end{table}

So far we have only discussed harmonic frequencies. The effect of
anharmonicity can be treated using either a Taylor expansion of the
PES in terms of normal mode coordinates or by explicitly spanning
the PES on a numerical grid. The discussion of anharmonic force
constants is postponed to the following section. Here, we will focus
on an explicit PES generated by means of the following correlation
expansion, here written up to three-mode correlations,
\cite{carter97:10458}
\begin{equation}
\label{eq:correxp}
V(\mathbf{ Q})=\sum_i V^{(1)}(Q_i)+\sum_{i<j}V^{(2)}(Q_i,Q_j)+\sum_{i<j<k}V^{(3)}(Q_i,Q_j,Q_k)\, .
\end{equation}
Neglecting rotation \cite{watson68:479}, i.e. assuming
that the kinetic energy operator is diagonal, the eigenstates of
the respective Hamiltonian can be obtained by straightforward
diagonalization using, e.g., the Lanczos method. For systems of the
size of solvated base pairs the calculation of higher order
correlation terms becomes rather expensive. Here, an interesting
alternative are so-called dual level schemes where low-order
correlation PES are calculated at a higher level of quantum
chemistry than multi-mode correlation PES. For instance, Scheurer
and coworker have combined MP2 and PM3 calculations to find a rather
good description of IR spectra of a model peptide
\cite{bounouar06:87}.

In Table \ref{tab2} we present results of dual level calculations on
4D models including the target modes of Figs. \ref{DNA:fig:NMs_1_2}
and \ref{DNA:fig:NMs_3_4}. Here the one-mode potentials,
$V^{(1)}(Q_i)$, have been calculated using the DFT method, while
two- and three mode PES were generated using the semiempirical PM3
approach.
\begin{table}[tdp]
\caption{Anharmonic frequencies (in \cm) for the target modes of the
different model structures as obtained using Eq. (\ref{eq:correxp})
with the one mode potential generated by the DFT/B3LYP method with a
6-31++G(d,p) basis set and the two- and three-mode PES obtained by
the PM3 approach.}
\begin{center}
\begin{tabular}{lcccc}
\hline
mode & AT-H$_2$O &AT-(H$_2$O)$_2$ &AT-(H$_2$O)$_3$ &AT-(H$_2$O)$_4$ \\
\hline
\nNH2 & 3326 & 3332 & 3307& 3310\\
\nCnHB & 1803 & 1813& 1797& 1785 \\
\dNH2 & 1732 & 1760& 1777& 1753\\
\nCHB & 1701 & 1664& 1635& 1643\\
\hline
\end{tabular}
\end{center}
\label{tab2}
\end{table}
\begin{table}[tdp]
\caption{Comparison of anharmonic frequencies (in \cm) for the
target modes of the AT-(H$_2$O)$_2$ model structures of different
two-dimensional calculations and different levels of quantum
chemistry (DFT: full DFT/B3LYP, 6-31++G(d,p); PM3: full PM3; DUAL:
DFT one-mode and PM3 two-mode PES). }
\begin{center}
\begin{tabular}{lcccc}
\hline
2D model & mode &DFT & PM3 & DUAL\\
\hline
\nNH2,\dNH2 &\nNH2 & 3311 & 3154 & 3321\\
& \dNH2 & 1727 & 1950 & 1730 \\
\hline
\nCnHB, \dNH2 & \nCnHB & 1796 & 1901 & 1796 \\
& \dNH2 & 1742 & 1703 & 1742 \\
\hline
\nCHB,\dNH2 & \nCHB &1715& 1807& 1680\\
& \dNH2 & 1744 & 1671 & 1777 \\
\hline
\end{tabular}
\end{center}
\label{tab3}
\end{table}
Comparing these anharmonic results with the harmonic values in Table
\ref{tab1} we notice the following: The \nNH2 vibration is strongly
affected and red-shifts by about 70-100 \cm depending on the cluster
size. The \nCnHB vibration is only slightly affected. Essentially,
these two modes behave as expected. Except for the AT-H$_2$O case
the \dNH2 vibration is blue-shifted by about 40 \cm. The fundamental
transition of the \nCHB mode, on the other hand, is considerably
red-shifted by about 30-80 \cm. In terms of the experimental
assignment given in Table 6 below the \nCHB agrees
fairly well with experiment which puts this transition at 1665 \cm
\cite{tsuboi69:45,clark85,howard65:801}. However, the \dNH2 vibration is believed to absorb around
1665 \cm as well, which is at variance with the prediction of the
dual level scheme. Moreover, in the harmonic case but also in the
fourth order anharmonic force field calculations reported below the
frequencies of \nCHB and \dNH2 are almost identical. It would be
surprising if the higher order anharmonic terms implicitly included
in the PES expansion changes this situation to such an extent.

Since multimode calculations are rather expensive we have chosen to
scrutinize the effect of the PM3 approximation by comparing
different 2D models of AT-(H$_2$O)$_2$ at the full DFT, PM3, and
dual level. The results are compiled in Table \ref{tab3}. The
correlation between the \nNH2 and \dNH2 modes is rather
well-desribed by the dual level scheme, the error being just a few
\cm. The same holds true for the correlation between the \nCnHB and
\dNH2 modes. However, the hybrid scheme is performing poorly for
the correlation between the \nCHB and \dNH2 modes and gives the
frequency shifts of opposite sign also observed for the full 4D
calculation in Table \ref{tab2}. The failure of the dual level scheme
to describe the coupling of the \nCHB and \dNH2 bending motion can
be understood in terms of their considerable mixing as quantified by a normal mode internal coordinate
decomposition. In general, an internal coordinate may contribute to
the displacement along several normal modes. Analyzing the present situation using the scheme of Boatz and
Gordon \cite{boatz89:1819} the internal coordinate describing    
the NH$_2$ angle contributes to the decomposition of the \dNH2 bending
normal mode by only 29 \%, while its contribution to the \nCHB
stretching normal mode (see also Fig. \ref{DNA:fig:NMs_1_2} ) and several purine ring
deformations normal modes ranges between 15 and 17\% . On the
other hand, the internal NH$_2$ angle does not contribute to the
decomposition of the \nCnHB and \nNH2 normal modes which are 92\%
stretching of the C2=O2 bond and 82\% and 17\% stretchings of the
N-H bonds, respectively.
%
\subsection{Anharmonic Coupling Patterns}
\label{sec:afc}
In this section we explore the second possibility to generate
multidimensional PES, i.e. a Taylor expansion in terms of normal
mode coordinate with respect to the geometry of the stable
structure. Including terms of to fourth order we have (using
dimensionless coordinates)
\begin{equation}
\label{eq:vafc}
V(\mathbf{Q})=\frac{1}{2!} \sum_i \hbar \omega_i Q_i^2+\frac{1}{3!}\sum_{ijk}K^{(3)}_{ijk}Q_iQ_jQ_k+\frac{1}{4!}\sum_{ijkl}K^{(4)}_{ijkl}Q_iQ_jQ_kQ_l \, .
\end{equation}
Third and fourth order anharmonic coupling constants are then
calculated using a combination of analytical second derivatives and
finite differences \cite{schneider89:367}. Specifically, we have
used the symmetric expressions \cite{csaszar98:13}
\begin{equation}
\label{ }
K^{(3)}_{ijk} = \frac{-K_{ij}^{(+2)} + 8 K_{ij}^{(+)} - 8 K_{ij}^{(-)} + K_{ij}^{(-2)}}{12 \Delta Q_k} \, ,
\end{equation}
\begin{equation}
\label{ }
K^{(4)}_{ijkl} =\frac{K_{ij}^{(++)} + K_{ij}^{(+-)} - K_{ij}^{(-+)} + K_{ij}^{(+-)}}{4 \Delta Q_k \Delta Q_l} \, .
\end{equation}
Here, $K_{ij}^{\pm}$ is the Hessian calculated at displaced
geometries where for the displacement we used $\Delta Q=0.03$ for
the cubic and $\Delta Q=0.04$ for the quartic force field. Note that
for the construction of the Hamiltonian we have neglected
contributions which are off-resonant by more than $\sim$1000 \cm.
\begin{table}[tdp]
\caption{Harmonic and diagonal anharmonic force constants (in \cm) of the relevant system modes.}
\begin{center}
\begin{tabular}{lccccc}
\hline 
& \nNH2 & \nCnHB & \dNH2 & \nCHB\\[1.0ex]
\hline
$\omega$ & 3401 & 1799 & 1713 & 1720 \\
$K^{(3)}/3!$ & -261 & 61 & -19 & -31\\
$K^{(4)}/4!$ & 24 & 5 & 4 & 2 \\
\hline 
\end{tabular}
\end{center}
\label{tab:K2}
\end{table}

In the following we will focus on the case of two water molecules
only, i.e. AT-(H$_2$O)$_2$, since this already contains the
essential effect of hydrogen-bonding waters as discussed in the previous section. The diagonal force
constants for the four target modes of Fig. \ref{DNA:fig:NMs_1_2}
(right panel) are given in Table 4. Important third order
anharmonic coupling constants involving the \nNH2 mode are compiled
in Table \ref{tab:K3}. As expected
the diagonal anharmonic force constants are largest for the \nNH2
mode. More interesting, however, is the coupling pattern between
this mode and the fingerprint modes. Here, we observe two dominating
Fermi-type resonance couplings: (i) to the bending overtone 2\dNH2
which is by far the strongest coupling, (ii) to the combination tone
between the \dNH2 and the \nCHB modes. In Table \ref{tab:K3} we also
give the couplings for the isolated AT case. Notice that here only
the 2\dNH2 overtone is strongly coupled. In other words, the
presence of water establishes as new coupling channel. And, going
back to Fig. \ref{DNA:fig:NMs_1_2} it requires a water molecule at
the C4=O4 site which mixes the \dNH2 and the \nCHB modes.
\begin{table}[tdp]
\caption{Third order coupling constants between the \nNH2 mode and
relevant fingerprint modes (in \cm) in AT-(H$_2$O)$_2$. The numbers
in parentheses refer to the isolated AT case.}
\begin{center}\begin{tabular}{lcccc}
\hline
& \nCnHB & \dNH2 & \nCHB \\
\hline
\nCnHB & -1 (-2) & 5 (8) & 0 (-2) \\
\dNH2 & & 96 (87) & -66 (-17) \\
\nCHB& & & 11 (2)\\
\hline
\end{tabular}
\end{center}
\label{tab:K3}
\end{table}
\begin{table}[tdp]
\begin{center}
\caption{Anharmonic frequencies (in \cm) for the target modes of the
AT-(H$_2$O)$_2$ model calculated using the anharmonic expansion, Eq.
(\ref{eq:vafc}) and the DFT/B3LYP method with a 6-31++G(d,p) basis
set. The 4D model is compared with a 6D model which includes in
addition the most strongly coupled water stretching and bending
modes at the N6-H site whose anharmonic frequencies are 3752 \cm and
1588 \cm, respectively. Also given are results for a 4D model which
does not include water molecules \cite{heyne08:7909}. The
experimental assignment is shown as well (\nNH2 from Ref.
\cite{heyne08:7909}, the other modes from Refs. \cite{tsuboi69:45,clark85,lettelier87:663,howard65:801,miles64:1104,adam86:3220,liquier91:177,sutherland57:446,ghomi90:691}).}
\begin{tabular}{lcccc}
\hline
mode & 4D (AT) & 4D & 6D & exp. \\
\hline
\nNH2 & 3330 &3297 & 3280 & 3215\\
\nCnHB & 1758 &1796 & 1792 & 1716 \\
\dNH2 & 1719 & 1718& 1708& 1665\\
\nCHB & 1645 & 1709& 1702& 1665\\
\hline
\end{tabular}
\end{center}
\label{tab:afcfreq}
\end{table}

The fundamental transition frequencies obtained from this 4D
anharmonic force field are given in Table 6 which
besides the experimental values contains frequencies calculated for
a 6D model which additionally includes most strongly coupled water
stretching and bending modes at the N6-H site \cite{heyne08:7909}.
Inspecting 4D and 6D cases we notice that the effect of explicit
inclusion of water modes is only modest especially in comparison to
the isolated case (4D(AT), see also Table \ref{tab1}). Given the
simplicity of the model, the agreement between theory and experiment
is rather reasonable with deviations being about 2\% except for the
\nCnHB mode whose frequency is about 4\% below the experimental
value. Perhaps this is not very surprising as the C2=O2 mode is
close to the thymine N1 site where in DNA the base is linked to the
backbone.

\section{Experimental Section}
\subsection{Methods}
\label{sec:exp}
A-T DNA double strand oligomers with sodium counterions and a length
of 20 base pairs were obtained from Biotherm, and were dissolved in
water and dried on a CaF$_{2}$ window at 293 K in an atmosphere of 52\%
relative humidity (saturated solution of NaHSO$_4$.H$_2$O at
20$^\circ$ Celsius \cite{obrien48:73}). This results in DNA samples with
approximately 4 to 6 water molecules per base pair \cite{sutherland57:446} (sample
thickness $\sim$ 6.5 $\mu$m). It has been reported that under these
conditions AT DNA oligomers adopt the B'-form \cite{adam86:3220}. Femtosecond
time-resolved IR pump-probe experiments were performed with
two independently tunable femtosecond pulses generated by parametric
conversion processes pumped by a regenerative Ti:sapphire laser
system (800 nm, repetition rate 1 kHz, pulse duration 100 fs) \cite{kaindl00:2086}.
The central frequency of the pump pulse was varied from 1630 to 1760
\cm and the probe was centred around 1650 \cm or 3200 \cm. The cross
correlation between pump and probe pulses had a temporal
width of ~130 fs (FWHM). With the used pump pulse energy of 1 $\mu$J
approximately 2\% of the AT base pairs in the sample volume were
excited. After interaction with the sample, the probe pulses were
spectrally dispersed and detected with a HgCdTe detector array
(resolution 5 \cm).
\subsection{Experimental Results}
\begin{figure}[t]
\begin{center}
\includegraphics[scale=0.8]{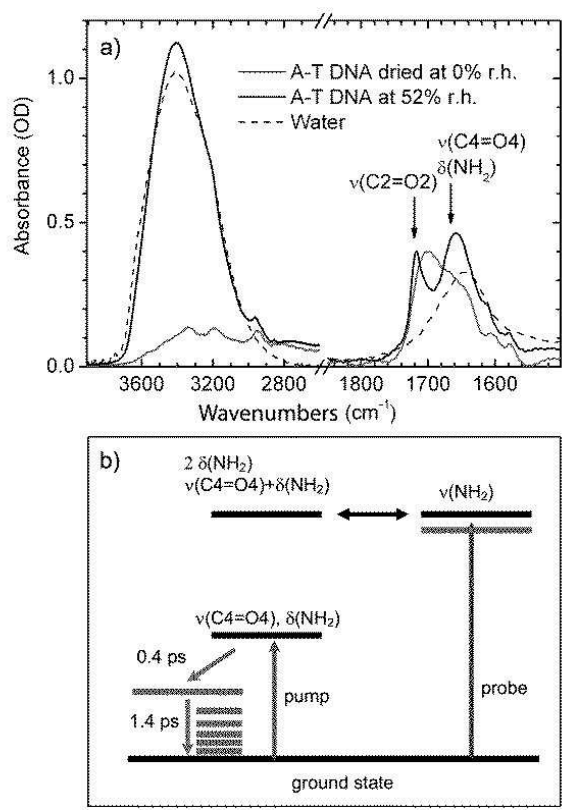}
\caption{(a) Absorption spectra of AT DNA oligomers prepared in
52\% relative humidity (black solid line), neat water (dashed line),
and AT DNA oligomers dried for two days in a N$_2$ atmosphere (gray
solid line. (b) Scheme describing the two color IR pump-probe
detection of the \nNH2 stretching vibration in AT base
pairs. Because of the anharmonic coupling between the
\nCHB, \dNH2, and \nNH2 vibrations,
the \nNH2=0$\rightarrow\,$1 transition is bleached upon
excitation of the \nCHB and \dNH2 modes.
Polupation of the excited state levels will locally heat the
molecule, inducing a shift of the hot ground-state \nNH2
transition.} \label{picmod}
\end{center}
\end{figure}

\begin{figure}[t]
\begin{center}
\includegraphics[scale=.75]{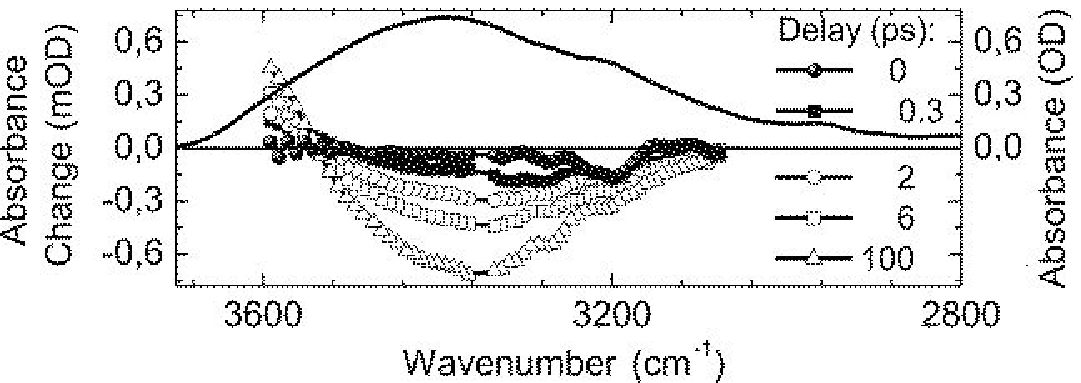}
\caption{Absorption spectrum of AT DNA oligomers around 3300 \cm
(solid line) and absorbance difference spectra for several
pump-probe delay times after excitation at 1740 \cm (FWHM 170 \cm).
The picosecond OH stretching response of water ranges from 3600 to 3050 \cm.
The spectrum at 0 ps delay time were obtained by averaging from -200
to 200 fs to eliminate nonabsorbing signal contributions.}
\label{pic5}
\end{center}
\end{figure}

The absorption between 3050 and 3600 \cm (see Fig. \ref{picmod}a and
\ref{pic5}, solid line) is dominated for more than 85\% by the broad
OH stretching absorption of water molecules. Therefore, it is not
possible to directly determine the \nNH2 stretch absorption
frequency of the DNA oligomer from this absorption spectrum. In the
fingerprint region, the absorption of \nCnHB is located at 1716 \cm
and the combined absorption of \dNH2 and \nCHB peaks at 1665 \cm
(see Fig. \ref{picmod}a, solid line) \cite{tsuboi69:45,clark85,howard65:801,falk63:387,ghomi90:691,krishnan07:132}. Both the \nCHB
and the \dNH2 vibration absorb at 1665 \cm and therefore
cannot be excited separately in our experiment. Figure \ref{pic5}
shows results of femtosecond pump-probe experiments with excitation
in the fingerprint region and probing between 3050 and 3600 \cm.
Excitation with a broad pump pulse at 1740 \cm (FWHM 170 \cm) leads
to an instantaneous spectrally narrow response around 3200 \cm.
Furthermore, a spectrally broad response over the entire range of
3050 \cm to 3600 \cm is seen to increase on a picosecond time scale
(see Fig. \ref{pic5}). At 1740 \cm the pump pulse mainly excites the
\nCnHB stretching vibration. Given the photon energy and pulse
intensity, two-photon excitation of vibrations around 3300 \cm is
unlikely by the pump pulse. As a consequence, both the instantaneous
and the increasing broad negative signal must result from exciting
vibrations in the fingerprint region. The broad negative signal,
which becomes positive above 3530 \cm, is known to correspond to the
OH stretching vibration of hot bulk water. Excess energy in low
frequency vibrations of water (e.g. librations) weakens the HB
strength, resulting in an increase of the OH stretching force
constant, and therefore a higher OH stretching frequency \cite{nibbering04:1887,nienhuys99:1494,huse05:389}.
The instantaneous narrow response around 3200 \cm, after excitation
at 1740 \cm, decreases in time and should therefore correspond to a
different process.
\begin{figure}[t]
\begin{center}
\includegraphics[scale=.5]{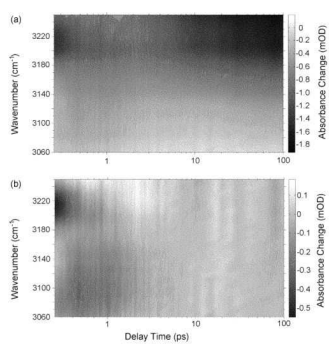}
\caption{(a) Contour plot of the absorbance change as a
function of delay time on a log scale after excitation at 1630 \cm
(FWHM 160 \cm). On the picosecond time scale the water response is
visible. (b) Contour plot of the absorbance change after
subtraction of the monoexponential 13 ps rise time of the hot water
response. The negative peak around 3215 \cm is clearly visible for
early delay times.} \label{pic1}
\end{center}
\end{figure}
Figure \ref{pic1} shows absorbance changes in the range between 3050
and 3250 \cm, upon excitation at 1630 \cm (FWHM 160 \cm), before and
after subtraction of the spectrally broad 13 ps component of the hot
water formation, obtained from a global fit. At 1630 \cm the pump
pulse mainly excites the \dNH2 and \nCHB
vibrations. An instantaneous bleach signal is observed with a
maximum at 3215 \cm and a width of 50 \cm, which decays on a
subpicosecond time scale. The perturbed free induction decay of this
band gives a total dephasing time T$_2$ of 0.5 $\pm$ 0.1 ps, which
corresponds to a homogeneous line width of 21 $\pm$ 5 \cm. This
indicates that the origin of the observed 50 \cm width of the
bleaching band is not caused by a single homogenously broadened
absorption line. The maximum of the instantaneous response at 3215
\cm decays with a 0.6 $\pm$ 0.2 ps time constant. Around 3130 \cm a
positive signature seems to be present for early delay times
evolving into a negative band with a rise time of about 0.4 ps, that
decays with a time constant of 1.4 $\pm$ 0.4 ps. The time constants characterizing the kinetics for
various pump-probe wavelength combinations are summarized in Table
\ref{timeconst}. 
\begin{table}[b]
\caption{Time constants of transients. r: rising signals; d:
decaying signals. }
\begin{center}
\begin{tabular}{lccc}
\hline
mode & pump (\cm) (FWHM) & probe (\cm) & time constants (ps) \\
\hline
\nNH2 & 1630 (160) & 3215 & (d) 0.6 $\pm$ 0.2 // (d) 3.0 $\pm$ 1.5 // (r) 13 $\pm$ 2\\
\nH2O$^a$ & 1630 (160) & 3130 & (r) 0.4 $\pm$ 0.2 // (d) 1.4 $\pm$ 0.4 // (r) 13 $\pm$ 2\\
\nNH2 & 1730 (90) & 3215 & (d) 0.9 $\pm$ 0.4 // (r) 4.0 $\pm$ 1.5 // (r) 13 \\
\nCnHB & 1760 (100) & 1725 & (d) 0.9 $\pm$ 0.1 \\
\nCnHB & 1760 (100) & 1685 & (d) 0.7 $\pm$ 0.1 \\
\nCnHB & 1630 (130) & 1720 & (d) 2.4 $\pm$ 0.2 \\
\nCHB / \dNH2 & 1630 (130) & 1665 & (d) 0.4 $\pm$ 0.1 // (d) 1.4 $\pm$ 0.4 \\
\dH2O & 1630 (130) & 1650 & (d) 0.2 $\pm$ 0.1 // (d) 1.0 $\pm$ 0.2 \\
\dH2O & 1630 (130) & 1640 & (r) 0.6 $\pm$ 0.1 \\
\nCHB / \dNH2 & 1630 (130) & 1625 & (d) 0.5 $\pm$ 0.1 \\
\hline
\end{tabular}
\end{center}
$^a$ Suggested assignment (see text)
\label{timeconst}
\end{table}

\begin{figure}[t]
\begin{center}
\includegraphics[scale=.5]{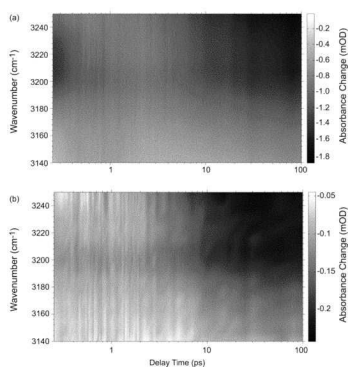}
\caption{(a) Contour plot of the absorbance change as a function of
delay time after excitation at 1630 \cm (FWHM 160 \cm) from 3140 to
3250 \cm. The instantaneous negative signal at 3215 \cm as well as
the dynamics of the hot water vibrations on a picosecond time scale
is clearly visible. (b) Contour plot of the absorbance change after
excitation at 1730 \cm (FWHM 90 \cm). The signals are much weaker.
The picosecond dynamics of hot water vibrations is clearly visible,
but the instantaneous response around 3215 \cm is negligible.}
\label{pic2}
\end{center}
\end{figure}

In order to identify the origin of the
instantaneous bleaching signal at 3215 \cm we compared transients at
3215 \cm after excitation at 1630 \cm and 1730 \cm, respectively, in Fig. \ref{pic2}.
Excitation at 1730 \cm results in a $\approx$ 7 times weaker and
barely visible instantaneous negative signal around 3215 \cm. In
contrast, excitation around 1630 \cm result in a pronounced
instantaneous signal around 3215 \cm. Comparing the two transients
at 3215 \cm, we observe that, excitation at 1630 \cm leads to a
three-exponential decay of the bleach with 0.6 ps, 3 ps, and 13 ps.
A transient with similar time constants (0.9 ps, 4 ps, and 13 ps),
but very different amplitudes is observed after pumping at 1730 \cm,
where mainly the \nCnHB stretching vibration is excited. Thus,
excitation in the spectral range of the \nCHB / \dNH2 vibrations
results in significant instantaneous signals around 3215 \cm.

\begin{figure}[t]
\begin{center}
\includegraphics[scale=.5]{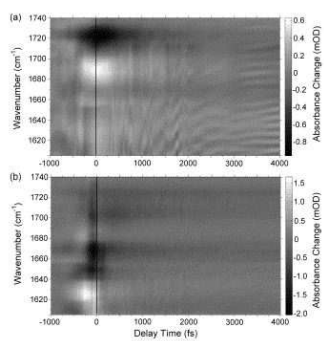}
\caption{(a) Contour plot of the absorbance change as a
function of delay time after excitation at 1760 \cm (FWHM 100 \cm).
(b)  Contour plot in the same frequency range (1605 to 1740
\cm) after excitation at 1630 \cm (FWHM 160 \cm).} \label{pic3}
\end{center}
\end{figure}

Experiments where both the pumping and probing takes place in the
fingerprint region are presented in Fig. \ref{pic3}. In Fig.
\ref{pic3}a the AT DNA oligomer sample was excited at 1760 \cm and
probed between 1605 and 1740 \cm. 
\begin{figure}[t]
\begin{center}
\includegraphics[scale=.45]{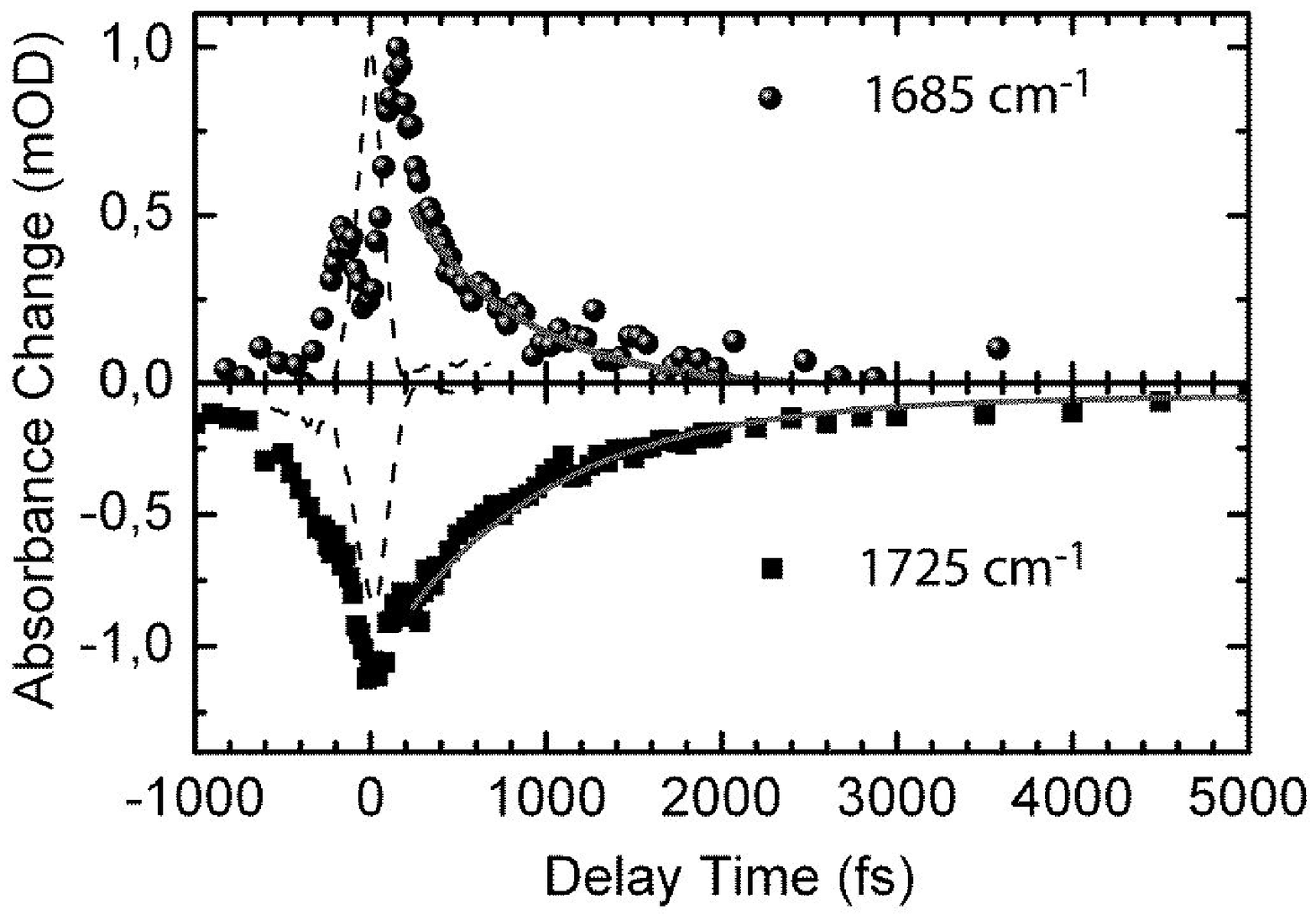}
\caption{Transients at 1685 \cm (circles) and 1725 \cm (squares)
after excitation at 1760 \cm (FWHM 100 \cm). Solid lines represent
fits with 0.7 $\pm$ 0.1 ps and 0.9 $\pm$ 0.1 ps time
constants for 1685 and 1725 \cm, respectively. System response
(dashed lines).} \label{picCO}
\end{center}
\end{figure}
The pump-probe spectrum shows a negative band at 1725 \cm and a positive band at
1685 \cm. The band at 1725 \cm decays mono-exponentially with 0.9
$\pm$ 0.1 ps, while the band at 1685 \cm decays with 0.7 $\pm$ 0.1
ps. Transients at these frequency positions are presented together
with their simulations in Fig. \ref{picCO}. The positive signal can
be assigned to the \nCnHB = 1 $\rightarrow$ 2 transition. A similar
lifetime was obtained by Zanni et al. for measurements on G-C DNA
oligomers with excitation in the same frequency region \cite{krummel03:9165}.  The
difference between the 0.7 ps excited state lifetime and the 0.9 ps
ground state recovery time signals is that the \nCnHB energy is
first converted into excitation of lower frequency modes. As a
consequence of this, the ground state absorption frequency is
shifted due to anharmonic coupling to these lower frequency modes,
and will not recover before these modes loose their excitation
energy. 

Results for excitation of the AT DNA oligomer at 1630 \cm,
presented in Fig. \ref{pic3}b, show bleaching signals at the ground
state absorption positions of the \nCnHB vibration (1716 \cm), the
\nCHB and \dNH2 vibrations (both 1665 \cm), and the water \dH2O
vibration (1650 \cm) \cite{tsuboi69:45,falk63:387,lee06:114508,huse05:389}.  The \nCnHB bleach recovers
exponentially with 2.4 $\pm$ 0.2 ps on a significantly longer time
scale than after direct \nCnHB excitation. This shows that there are
at least two different energy relaxation pathways in DNA involving
the \nCnHB vibration. For the \nCHB / \dNH2 vibrations
biexponential recoveries were observed with 0.4 $\pm$ 0.1 ps and 1.4
$\pm$ 0.4 ps, and for the water \dH2O with 0.2 $\pm$ 0.1 ps and 1.0
$\pm$ 0.2 ps. For the bending vibration \dH2O of water molecules in
bulk water a lifetime of 170 $\pm$ 30 fs \cite{huse05:389} has been reported, which
agrees with the fast component observed here for \dH2O.
Instantaneous increased absorption signals below 1640 \cm, decay
with a time constant of 0.5 $\pm$ 0.1 ps, matching the fast
component of the \nCHB / \dNH2 bleach recovery. In addition, a
positive band appears around 1640 \cm, rising in 0.6 $\pm$ 0.1 ps.
Signals at this spectral position have been assigned to the bending
vibration of hot water molecules \cite{huse05:389}.

The dynamics of experiments with different pump frequencies in the
fingerprint region, are compared to confirm that the bleach band at
3215 \cm originates from the \nNH2 stretching vibration of adenine
and not from water. First, we point out, that even the fastest
recovery time of the bleach signal at 3215 \cm after excitation in
the fingerprint region (where the water \dH2O also absorbs), is
three times slower than the reported lifetime of the water \dH2O
vibration \cite{huse05:389}. Second, the width of the bleaching band at 3215 \cm of
50 \cm is considerably narrower than the 200 \cm estimate for the
\dH2O overtone band of bulk water \cite{wang04:9054}.  Having assigned the 3215 \cm
bleaching signal to the AT DNA oligomer \nNH2 vibration, we now
compare the dynamics at this frequency for excitation at 1630 \cm
vibration and 1730 \cm, which correspond to the absorption bands of
both the \nCHB and \dNH2 vibrations, and the \nCnHB vibration,
respectively.

For both excitation frequencies the picosecond dynamics, shown in
Fig. \ref{pic2}, can be modelled by the same 13 ps time constant,
corresponding to the rise of hot water signal due to energy transfer
from the DNA oligomer to the water molecules. However, the initial
sub-picosecond dynamics are markedly different for these two
excitation frequencies, further confirming that both signals can not
be due to the overtone excitation of the water \dNH2. Although the
absorption at both frequencies is comparable, the signal strength
after excitation at 1630 \cm is in fact seven times stronger than
after excitation at 1730 \cm, which indicates that the \nCnHB
vibration couples substantially weaker to the \nNH2 vibration than
both the \nCHB and the \dNH2 vibration.

\section{Discussion}
\label{sec:disc}

A model summarizing the theoretical and experimental results is given
 in Fig. \ref{picmod}b. The calculated couplings of the
\nCHB, \dNH2, and \nCnHB to the \nNH2 vibration indicate that
excitation of either of these modes should result in a bleaching
signal due to the shifting of the \nNH2 = 0 $\rightarrow$ 1
transition. The bleach signals in Fig. \ref{pic2} around 3215 \cm
agree with this theoretical result. Furthermore, the force constants
in Table \ref{tab:K3} predict that excitation of the \nCHB and \dNH2
vibrations should result in a larger shift of the \nNH2 vibration
than of the \nCnHB vibration. This is confirmed by the 
data in Fig. \ref{pic2}. Experimentally, one cannot distinguish
between the contributions of the \nCHB and \dNH2 modes. From the
force constants in Table \ref{tab:K3}, however, we conclude that
excitation of the \dNH2 vibration is expected to have the most
substantial effect on the shift of the \nNH2 vibration. The
agreement between theoretical predictions and IR pump-probe
measurements allows us to assign the bleaching signal at 3215 \cm at
least partially to the symmetric \nNH2 vibration of adenine. This absorption
band lies about 100 \cm lower in energy than the same mode in
modified adenine-uracil Watson-Crick base pairs in solution \cite{woutersen04:5381}. The
lower frequency of the hydrogen-bonded \nNH2 vibration in DNA films
compared to single AU base pairs in CDCl$_3$ solution can be
rationalized by significant interactions with neighbouring base
pairs and water molecules, that weaken the force constant. The
theoretical results further show, that inclusion of water molecules
leads to a HB between a water molecule and the NH$_2$
group of adenine, and therefore a coupling of the adenine \nNH2
vibration and of the water bending vibration. Due to this coupling,
this water molecule could act as a primary energy sink in energy
disposal by DNA. Since the amount of water molecules and the
coupling of water molecules to the nucleic acids is different for
the major groove and the minor groove, one would expect different
energy relaxation pathways with deviating time constants for energy
flow from water to DNA and DNA to water, for both sides. Energy
redistribution processes in DNA itself provide complex relaxation
patterns as presented for the \nCnHB vibration.

In summary, the presented results demonstrate the capacity of
combining IR-pump-probe methods with calculations on microsolvated base pairs to reveal
information on and identify hidden vibrational absorption bands. The simulation of real condensed phase dynamics of HBs, however, 
requires to take into account all intra- and intermolecular interactions mentioned in the Introduction.  As far as DNA is concerned, Cho and coworkers have given an impressive account on the dynamics of the CO fingerprint modes \cite{lee06:114508,lee06:114509,lee06:114510,lee07:145102}. Promising results for a single AU pair in deuterochloroform \cite{woutersen04:5381} have been reported recently using a Car-Parrinello based QM/MM scheme \cite{yanxx}.
\begin{acknowledgement}
We gratefully acknowledge financial support by the Deutsche
Forschungsgemeinschaft (Sfb450) and the project MZOS
098-0352851-2921. We thank G. Kovacevic for his help concerning the
normal mode decomposition.
\end{acknowledgement}


\end{document}